\documentclass[pre,twocolumn,floatfix,showpacs,showkeys]{revtex4}
\usepackage{hyperref}
\usepackage{amsmath}
\usepackage{amssymb}
\usepackage{amsfonts}
\usepackage{epsfig}
\usepackage{aecompl}
\usepackage{ae}

\newcommand{\Fig}[1]{Fig.~\ref{fig:#1}}

\def\bmath#1{\mbox{\boldmath$#1$}}

\begin{document}

\title{
  Morphologies of three-dimensional shear bands
  in granular media
}

\author{
  S. Fazekas$^{1,2}$, J. T\"or\"ok$^{2,3}$, J. Kert\'esz$^{2}$,
  and D. E. Wolf$^{4}$
}

\affiliation{
  $^1$Theoretical Solid State Research Group of
      the Hungarian Academy of Sciences,\\
  $^2$Department of Theoretical Physics, and \\
  $^3$Department of Chemical Information Technology,\\
      Budapest University of Technology and Economics,\\
      H-1111 Budapest, Hungary\\
  $^4$Department of Physics, University Duisburg-Essen,\\
      D-47048 Duisburg, Germany
}

\date{February 20, 2006}

\begin{abstract}
  We present numerical results on spontaneous symmetry breaking strain
  localization in axisymmetric triaxial shear tests of granular materials.
  We simulated shear band formation using three-dimensional
  Distinct Element Method with spherical particles. We demonstrate that
  the local shear intensity, the angular velocity of the grains,
  the coordination number, and the local void ratio are correlated and
  any of them can be used to identify shear bands, however the latter
  two are less sensitive. The calculated shear band morphologies
  are in good agreement with those found experimentally. We show that
  boundary conditions play an important role. We discuss the formation
  mechanism of shear bands in the light of our observations and compare
  the results with experiments. At large strains, with enforced symmetry,
  we found strain hardening.
\end{abstract}



\maketitle

\section{Introduction}

The description of the rheological properties of dry granular media is
a key question which controls the ability of handling (mixing, storing,
transporting, etc.) of these particulate systems. An interesting and
sometimes annoying feature of such materials is strain localization
which appears almost always when a sample is subjected to deformation.
The morphology of these narrow domains (shear bands) is far from being
understood.

Two-dimensional and boundary induced shear band shapes have a vast
literature dating back to decades including both numerical and
experimental studies. Three-dimensional studies have a major drawback
in the difficulty of getting information from inside the sample.
However, in the
past 20 years they gained increasing attention as experimental tools
as Computer Tomography (CT) became available. Such experimental
studies \cite{desrues-geo96,desrues-ct04,alshibli-gtj00,batiste-gtj04}
revealed complex localization patterns and shear band
morphologies depending on the test conditions.

In this paper we focus on triaxial tests, which in general
are elementary tests, performed to obtain mechanical
properties of soils.
The most common axisymmetric triaxial test consists of a cylindrical
specimen enclosed between two end platens and surrounded by a rubber
membrane on which an external pressure is applied
(see for example \cite{desrues-ct04}). The end platens are pressed
against each other in a controlled way: Either with constant velocity
(strain control) or with constant force (stress control). The force
resulting on the platens or the displacement rate of the platens is
recorded as well as the volume change of the specimen.

We report a numerical study of triaxial tests of
cohesionless granular materials based on three-dimensional
Distinct Element Method (DEM). We show that depending on the boundary
conditions different shear band morphologies can be observed similar
to experiments. We identify the shear bands
by calculating the local shear intensity and show that it is
correlated with the angular velocity of the particles and also
with the local void ratio and the coordination number, which give
alternative ways to detect shear bands and further verification
of our results.

\section{Simulations}

\begin{figure}[t!]
\begin{center}
\begin{tabular}{ccc}
\epsfig{file=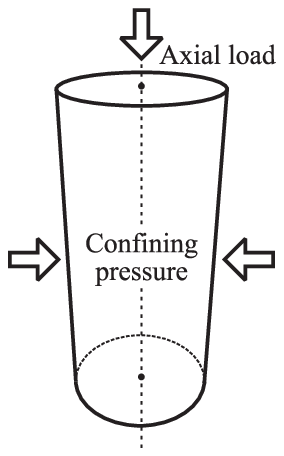,height=4cm}&
\epsfig{file=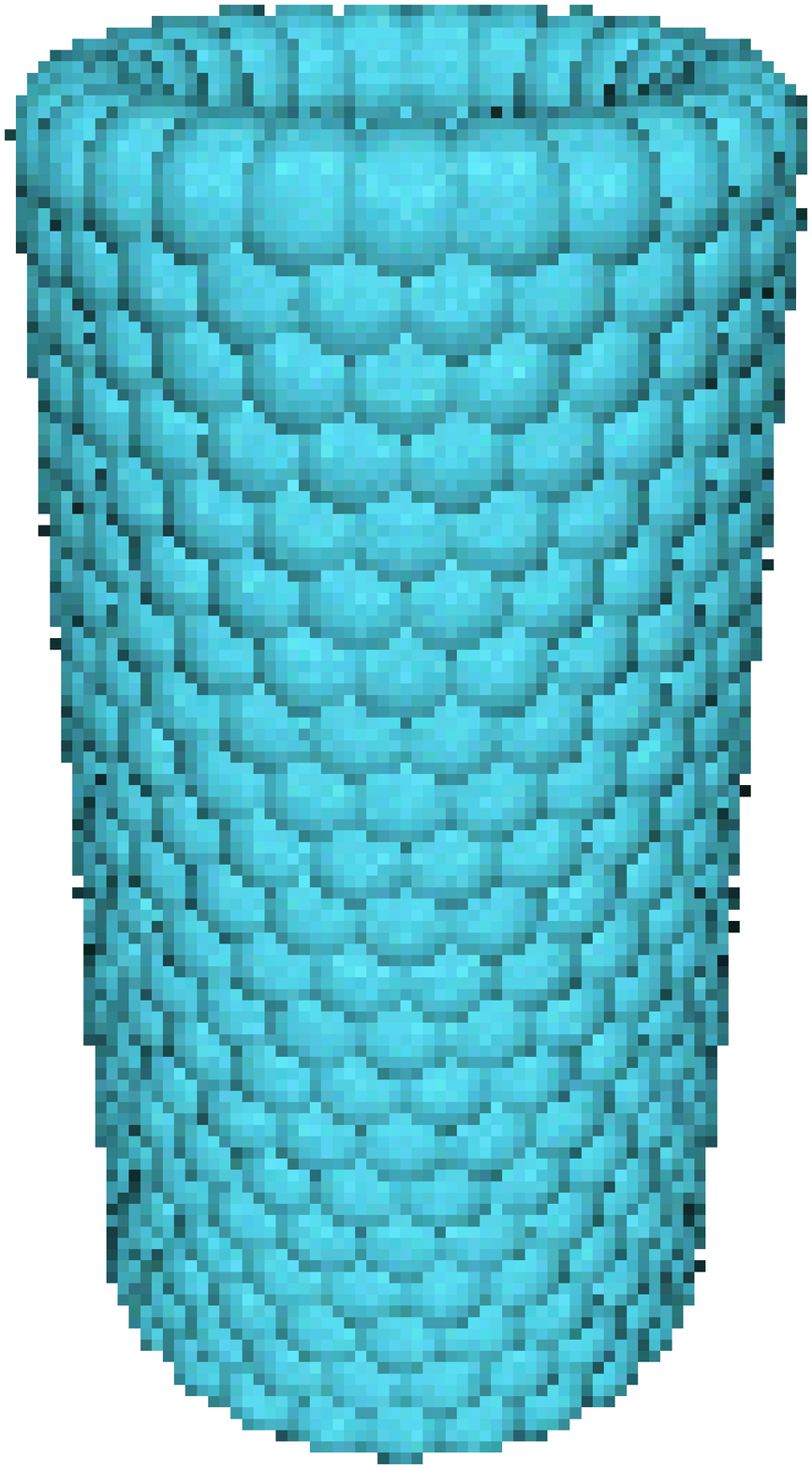,height=4cm}&
\epsfig{file=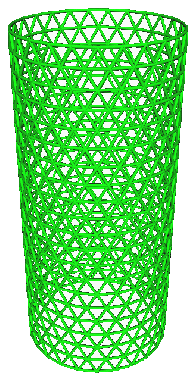,height=4cm}\cr
a)&b)&c)
\end{tabular}
\end{center}
\caption{
  (Color online)
  a) A granular sample was subjected to axial load and
  confining pressure. b) The rubber membrane surrounding the
  sample was simulated by overlapping spheres initially arranged in a
  triangular lattice. c) The neighboring spheres were interconnected
  with linear springs. The confining pressure acted on the
  triangular facets.
}
\label{fig:simmod}
\end{figure}

We used a standard DEM with
Hertz contact model \cite{landau-70} and appropriate damping
\cite{brilliantov-pre96} combined with a frictional spring-dashpot
model \cite{luding-book04,poeschel-05}.
The triaxial tests were performed on vertical cylindrical samples
(see \Fig{simmod}) of diameter $D=22\ mm$ and height $H=46\ mm$
(i.e. the slenderness was $H/D \approx 2.1$ similar to most experiments).
Each sample consisted of $27000$ spherical particles with the same
mass density $\varrho=7.5\cdot 10^3\ kg/m^3$.
The particles had a Gaussian size distribution. The mean particle
diameter was $d =0.9\ mm$. The standard deviation of the
particle diameters was $\Delta d=0.025\ mm$.

The normal $F_n$ and tangential ${\bf F}_t$ components
of the contact force were calculated as
\begin{eqnarray}
  F_n &=&
    {\kappa}_n {\delta}_n^{3/2}
    - {\gamma}_n {\delta}_n^{1/2} {v_n}, \\
  {\bf F}_t &=&
    {\kappa}_t {\bmath{\delta}}_t
    - {\gamma}_t {\bf v}_t,
\end{eqnarray}
where ${\kappa}_{n}=10^{6}\ N/m^{3/2}$,
${\kappa}_{t}=10^{4}\ N/m$, ${\gamma}_{n}=1\ N\:s/m^{3/2}$, and
${\gamma}_{t}=1\ N\:s/m$ are the normal and tangential stiffness and
damping coefficients, ${\delta}_n$ and ${\bmath{\delta}}_t$ are the
normal and tangential displacements, and ${v}_n$ and ${\bf v}_t$ are
the normal and tangential relative velocities.

The normal displacement was calculated from the relative position,
the size, and the shape of the bodies in contact.
The tangential displacement was calculated
by integrating the tangential velocity
in the contact plane during the lifetime of the contact.
The Coulomb law limits the (tangential)
friction force to $\mu F_n$ (where $\mu=0.5$
is the used coefficient of friction). To allow for sliding contacts,
we limited the length of the tangential displacement to
$\mu F_n / {\kappa}_t$.
(For a review on DEM see \cite{luding-book04,poeschel-05} and
references therein. For more details on our implementation see
\cite{fazekas-png05}.)

The translational motion of bodies is calculated with Verlet's
leap-frog method. The rotational state
is integrated in quaternion representation with Euler's
method. With the above stiffness and
damping coefficients, the inverse of the average eigenfrequency of
contacts, in both normal and tangential direction, is more
than one order of magnitude larger than the used integration time step
$\Delta t=10^{-6}\ s$. This assures that the noise level induced
by numerical errors and grain elasticity is kept low.

The initial configuration was generated by randomly placing
the grains in a tall solid cylinder having height $h=3H$ and
width $D$. The maximum allowed initial grain overlap was $1\%$.
The upper platen and the particles
were given downward velocities $v \varpropto z/h$
depending on their vertical position $z$ measured
from the fixed bottom platen. These conditions lead to an
almost simultaneous first contact of the bodies.
The system was stabilized with a force applied on the upper platen.
This was switched on when the inner pressure of the sample
could compensate it.
In preparation (and later in the tests) we used zero gravity.
The deposition method described
above is known to produce a homogeneous system.
During the preparation phase, the friction of the
particles was switched off to
allow for generation of dense samples.
(For a review on sphere packings see \cite{weitz-sci04}.)

The solid cylinder used in
preparation was replaced by an \emph{elastic membrane}
in the tests.
The elastic membrane was modeled with overlapping
spheres having equal diameter $d_m = 1\ mm$ and equal mass density
$\varrho_m = 100\ kg/m^3$, and initially forming a triangular
lattice on the external surface of the cylinder. The ``membrane nodes''
could not rotate and were interconnected with linear springs
having an elongation equal to the relative distance of the nodes
(initially $0.5\ mm$). The stiffness of the
springs $\kappa_s=0.5\ N/m$ was chosen such that the particles
could not escape by passing through the membrane. Additionally a
homogeneous confining pressure ${\sigma}_c=500\ N/m^2$ was applied on
the membrane. This was simulated by calculating the forces acting
on the triangular facets formed by connected membrane nodes.

A similar model was used by Tsunekawa and Iwashita
\cite{tsunekawa-pg01} who applied the confining pressure directly on
the external particles in a very similar way. However, their
approach requires the computationally expensive identification of
external particles and a Delaunay triangulation. Sakaguchi and M\"uhlhaus
\cite{sakaguchi-eg01} used a similar membrane model to ours but without
an explicit confining pressure, relying only on the stiffness of the
springs.

The bottom platen was fixed during both preparation and
test phases. The upper platen could not tilt in preparation phase,
but in certain tests it could freely tilt along any horizontal axis
with rotational inertia $I=10^{-7}\ kg\:m^2$.
During the tests, the samples were compressed by moving the
upper platen in vertical direction downwards with a constant velocity
(strain control). Starting from the same initial condition, four different
runs -- denoted by (A), (B), (C), and (D) -- were executed.
Two different compression velocities were used: A base value
$u_1=10\ mm/s$ (in tests (A) and (B)) and a two times larger value
$u_2=2 u_1$ (in tests (C) and (D)).
Tilting of the upper platen was enabled
in tests (A) and (C) and disabled in tests (B) and (D).

\section{Results}

\subsection{Local shear intensity}

We define the local shear intensity $S$ by generalizing its
two-dimensional definition given by Daudon et al.
\cite{daudon-pg97}. First, the
regular triangulation \cite{lee-book91} of the particle system is
calculated \cite{CGAL-2.4}. The displacements of the particles
(relative to a previous state) are known from the DEM simulation.
We extend the displacement field to the whole volume of the sample
with a linear interpolation over the tetrahedra of the regular
triangulation. For each particle we identify the incident
triangulation cells (tetrahedra) which define a discrete particle
neighborhood. The particles at the sample's boundary, having infinite
incident cells, are skipped (i.e. no local shear intensity is defined
for them). The discrete neighborhood of a particle is a surrounding
polyhedron with first-neighbor particles at the corners.

We define the deformation gradient tensor with the partial (space)
derivatives $\partial_i u_j$ of the displacement vector $\mathbf{u}$.
In a neighborhood $\varOmega$ of volume $V$, the components of the mean
deformation gradient tensor are calculated as
\begin{equation}
  \langle u_{ij}\rangle
    = \frac{1}{V}\int_{\varOmega} \partial_i u_j\,dV.
\end{equation}
Using the Gauss-Ostrogradski theorem the volume integral
can be transformed into a closed surface integral over the boundary
$\partial\varOmega$ of $\varOmega$, leading to
\begin{equation} \langle u_{ij}\rangle =
\frac{1}{V}\oint_{\partial\varOmega} n_i u_j\,dS,
\end{equation}
where ${\bf n}$ is the exterior normal along the boundary.
We calculate the local deformation gradient tensor applying
the above formula to discrete neighborhoods and using the
linear interpolation of the particle displacements. In this
case the integral can be reduced to a summation over
triangular facets.

The symmetric part of the local deformation gradient tensor
is a macroscopic strain tensor derived from particle
displacements. Using the eigenvalues $\varepsilon_k$ of this macroscopic
strain tensor, we define the local shear intensity as
\begin{equation}
  S = \max_{k} \left| \varepsilon_k
      - \frac{1}{3} \sum_{l} \varepsilon_l \right|.
\end{equation}

We note that we disregard the elastic deformation and rotation of the
grains, since
we are interested in the identification of the shear bands, which is
strongly linked to geometric effects. However, for constitutive models,
a more complete treatment of the strain would be needed
\cite{osullivan-mg03}.

\subsection{Shear band morphologies}

\begin{figure}[t!]
\begin{tabular}{cc}
\epsfig{file=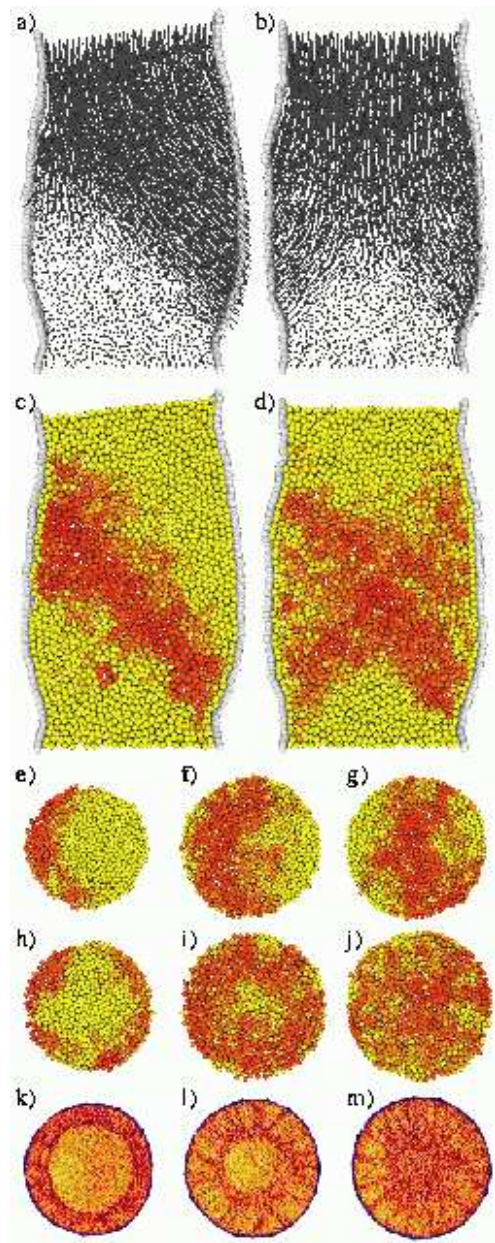,width=6.5cm}
\end{tabular}
\caption{
  (Color online)
  Cross sections (a, c, e, f, g) of sample (C) and
  cross sections (b, d, h, i, j) of sample (D)
  shown at $10\%$ axial strain. Panels (k, l, m) present CT scans
  \cite{batiste-gtj04} (F2075).
  The vertical cross sections (a-d) were taken at the middle
  of the sample. The horizontal cross sections
  were taken at different heights:
  close to the top (e, h, k), at the middle (g, j, m),
  and in between (f, i, l). The red color encodes
  the local shear intensity on panels (c-j) and
  the local void ratio on panels (k-m).
  Panels (a, b) show the velocity field.
}
\label{fig:cuts}
\end{figure}

Taking cross sections of the sheared samples
and coloring the grains according to the local shear intensity $S$,
we could identify shear bands (see \Fig{cuts}) and compared them with
experiments. In experiments the CT scans show the volume fraction
difference between the bulk and the shear band. In the next section we
justify the comparison of the volume fraction and local shear
intensity.

Our simulations are run at zero gravity and low confining pressure in very
similar conditions to the experiments of Batiste et al.
\cite{batiste-gtj04}. The shear band patterns found in their experiments
and our simulations are also very close to those found in experiments under
normal gravity and high confining pressure by Desrues et al.
\cite{desrues-ct04}, who also studied the case of a tilting upper platen.
We compared our results to both experiments.

We found that the absence of enforced axisymmetry leads to spontaneous
symmetry breaking. When tilting of the upper platen is enabled
internal instabilities can develop into a localized deformation along
a failure plane (see panel (c) of \Fig{cuts}).
Non-tilting platens act as a stabilizing factor leading
to an axisymmetric hourglass shaped shear band with two conical surfaces
and complex localization patterns around them
(see panel (d) of \Fig{cuts}).
This is in full accordance with the experimental results of
Desrues et al. \cite{desrues-ct04}.
For the non-tilting case Tsunekawa and Iwashita \cite{tsunekawa-pg01}
found in DEM simulations similar localization patterns, however they
have not investigated the tilting case.

In the non-tilting case, the horizontal cross sections (h, i, j)
shown on \Fig{cuts} can be compared with the experimental results
of Batiste et al. \cite{batiste-gtj04}.
They reported the same type of shear band morphologies
for these boundary conditions. The panels (k, l, m) of
\Fig{cuts} show CT scans from their triaxial shear tests executed
in micro-gravity aboard a NASA Space Shuttle. Our simulations used
similar setup and similar confining pressure. Good agreement of
shear band shapes (including their non-trivial structure)
can be recognized in spite of the rather limited number of grains
in our simulations. (Note that the details reproduce better in the
color version of the figure.)

In the case of tilting upper platen the shear bands are not totally
plane -- as can be seen on horizontal cross sections taken close to
the platens -- but follow the curvature of the boundary.
The same was found experimentally by
Desrues et al. \cite{desrues-ct04}.
In the vertical cross sections shown on the panels (c) and (d) of
\Fig{cuts},
the found shear bands are in good agreement with changes in the velocity
field shown on panels (a) and (b). This justifies the shear band
identification method based on the local shear intensity.
We have also investigated alternative methods.

\subsection{Alternative methods of shear band identification}

It is widely known that dense granular materials dilate during shear.
In some experiments
(e.g. experiments based on CT \cite{desrues-ct04,batiste-gtj04})
the local void ratio is used to identify the shear bands.
To confirm the presented shear band
identification method and the found shear band morphologies, we
have investigated the correlation between
the local void ratio $\nu$ and the local shear intensity $S$.
The void ratio was measured using the regular triangulation
\cite{lee-book91,CGAL-2.4} of the spherical particles.
The volume of the regular Voronoi cells $V_c$ and
the volume of the grains $V_g$
define the local void ratio $\nu=(V_c-V_g)/V_g$.

In numerical simulations (for spheres of nearly equal size),
a good alternative to the local void ratio $\nu$ is the coordination
number $Z$ (defined by the number of contacts), which decreases as
$\nu$ increases. Its main advantage is that it can be
defined exactly and calculated fast, however, if the size distribution is
wide a non-trivial particle size scaling has to be taken into account.

The existence of particle rotations in shear bands is known to
experimentalists for a long time (see e.g. \cite{oda-geo98}).
It was also evidenced in simulations by Herrmann et al.
\cite{herrmann-pha04}. In our simulations, we have also measured
for each grain the absolute value of the angular velocity $R$
and tested its correlation with the local shear intensity $S$.

\begin{figure}[t!]
\begin{tabular}{c}
\epsfig{file=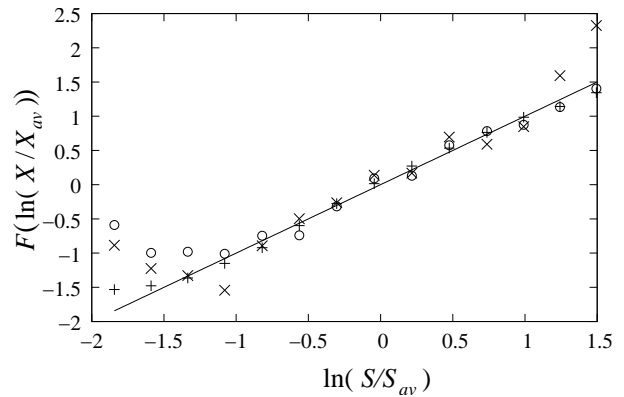,width=8cm} \\
\end{tabular}
\caption{
  Correlation of the local shear intensity $S$ with
  the local void ratio ($\nu$, $\circ$),
  the coordination number ($Z$, $\times$), and
  the angular velocity of the grains ($R$, $+$).
  $X$ is one of $\nu$, $Z$, and $R$.
  $F$ denotes a linear transformation different for each data set.
  All quantities are scaled by average values ($X_{av}, S_{av}$).
  The data is collected from four samples at $10\%$
  axial strain. See text for more details.
}
\label{fig:corr}
\end{figure}

All the quantities mentioned above
(the local void ratio $\nu$, the coordination number $Z$,
the angular velocity $R$, and the local shear
intensity $S$) are defined for each particle.
We checked their correlation with a histogram technique using
the values calculated for different particles
as different statistical samples.
The $\nu$, $Z$, and $R$ values were averaged for
each (logarithmic) histogram bin of $S$.
We also calculated the total averages
$\nu_{av}$, $Z_{av}$, $R_{av}$, and $S_{av}$.
On the quantities $\xi=\ln(X/X_{av})$ (where $X$ is one of
$\nu$, $Z$, and $R$) we applied different linear transformations
$F(\xi)=\alpha(\xi-\xi_0)$ (shift and scaling)
to achieve data collapse of $F(\ln(X/X_{av}))$ as
function of $\ln(S/S_{av})$ (see \Fig{corr}).

The scaling term $\alpha$ of $F$ shows the sensitivity of the
$R$, $Z$, and $\nu$ quantities with respect to the local shear
intensity $S$. We found $\alpha=1$ for the angular velocity,
$\alpha=-9$ for the coordination number, and $\alpha=27$ for the
local void ratio. (Note that $Z$ decreases as $S$ increases!)
The fluctuations were proportional to $\sqrt{|\alpha|}$.
Regarding shear band identification,
this means that the angular velocity is essentially
equivalent with the local shear intensity. However, the
coordination number and the local void ratio are less sensitive,
and they exhibit large fluctuations due to random packing
and random rearrangements.
For this reason they need more spatial and/or temporal averaging
to achieve the same accuracy.

\subsection{Stress-strain relation}

\begin{figure}[t!]
\epsfig{file=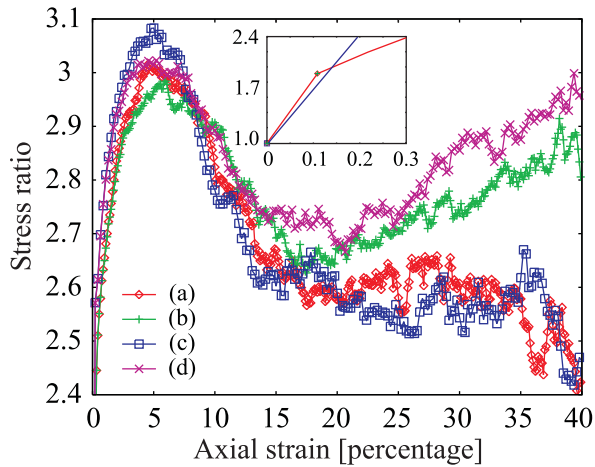}
\caption{
  (Color online)
  Stress-strain relation. The stress ratio $\sigma/\sigma_0$
  (where $\sigma_0$ denotes the initial stress)
  measured on the upper platen is shown as function of the axial strain,
  for different simulation runs. (See inset for low strains.)
  For the lower two curves (a, c) tilting of the upper platen was enabled,
  and for the upper two (b, d) it was disabled.
  For (c, d) the samples were  compressed two times faster
  than for (a, b).
}
\label{fig:strainfunc}
\end{figure}

In order to compare to most common experimental results,
we measured the stress $\sigma$ on the upper platen, and calculated
the stress ratio $\sigma/\sigma_0$, where $\sigma_0$ denotes the
initial stress. As the axial strain increases, the response of the
granular sample (the stress ratio) increases until it reaches a peak
value, then it decreases (see \Fig{strainfunc}). According to
\Fig{strainfunc}, up to $15\%$ axial strain there is no significant
difference in stress-strain relation measured in different
simulation runs, indifferent to strain rate and tilting of the
upper platen.
(We have not tested dependence on material
parameters and confining pressure.)

\begin{figure}[t!]
\epsfig{file=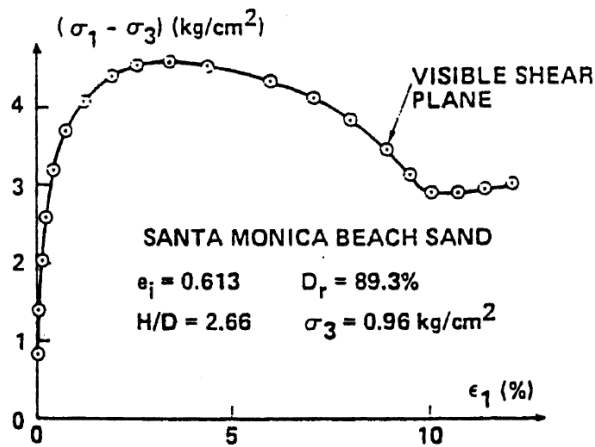,width=8cm}
\caption{
  Experimental stress-strain relation from a triaxial compression test
  exhibiting strain softening and development of shear plane.
  (Courtesy of Poul V. Lade, Reprinted from \cite{lade-ijss02} with
  permission from Elsevier.)
}
\label{fig:smbeachstress}
\end{figure}

The presented strain softening effect is a basic
observation of triaxial shear tests of dense granular specimens
(see for example \cite{lade-ijss02} and \Fig{smbeachstress}).
Any local deformation due to shear is followed
by a dilatation resulting in a decrease of the force bearing capacity of
the material which further intensifies the deformation leading to failure.
In our simulations we observed pronounced shear bands around $10\%$
axial strain, which is in good agreement with the experimental results.

After about $15\%$ axial strain the different boundary conditions
and shear band shapes result in different stress-strain curves.
In the non-tilting case, the geometry and the hourglass-shaped shear band
forces the particles to enter and leave the failure zone.
In this case, a stable slipping mode cannot be formed.
As the test sample is further compressed,
it opposes more and more firmly to compression
(see curves (b) and (d) on \Fig{strainfunc}),
resulting in increasing stress ratio (i.e. strain hardening).
In the tilting case, the upper part of the sample moves as
a single block. The formed planar shear band
allows for a stable slipping mode with
nearly constant stress  until boundary effects come into play
(see curves (a) and (c) on \Fig{strainfunc}).

\section{Conclusions}

Triaxial shear test simulations based on DEM were executed
and different shear band morphologies known from experiments were
reproduced
\footnote{We have ignored the rolling resistance
(see e.g. \cite{jiang-cg05}) in these
calculations. Our simulations demonstrate that this effect is not
crutial in the qualitative description of triaxial shear tests.}.
We have shown that in triaxial shear tests symmetry
breaking strain localization can develop spontaneously if the axial
symmetry is not enforced by non-tilting platens. To our knowledge
it is the first time that such symmetry breaking strain localization
was reproduced in DEM simulations.

We generalized the shear intensity definition of Daudon et al.
\cite{daudon-pg97} to three-dimensions and used it to identify
shear bands. To be able to compare to experiments,
we have also tested alternative methods of shear band
identification. We found strong
correlation of the local shear intensity with the angular velocity of
the grains, the coordination number, and the local void ratio.
This result justifies our method and proves once more the known
experimental and numerical findings that shear bands are characterized
by dilation and rotation of the grains.
Regarding shear band identification, the coordination number and
the local void ratio are found to be less sensitive
than the local shear intensity and the angular velocity of
the grains.

We have also measured the stress-strain relation of the compressed
samples. Strain softening was identified in good agreement with
experimental results. We have also found a strain hardening effect
at large strains in the non-tilting case and explained it in terms
of geometry and shear band morphology.
However, this might be only valid for the tested material
parameters and confining pressure.
We have no knowledge of experiments focusing on this particular question.
In general, the agreement of our results with the experimental results is
very good, even if the system size (number of particles) in our
simulations is much smaller than in experiments.

\section{Acknowledgments}

This research was carried out within the framework of the
``Center for Applied Mathematics and Computational Physics'' of the
BUTE, and it was supported by BMBF, grant HUN 02/011, and Hungarian
Grant OTKA T035028, F047259.

\bibliography{axsymtri}

\end{document}